\documentclass[%
 preprint,
 aip,
 jcp,
amsmath,amssymb,
longbiblioography,
]{revtex4-2}

\usepackage{graphicx}
\usepackage{dcolumn}
\usepackage{bm}
\usepackage{longtable}
\usepackage{verbatim}
\usepackage{setspace}
\usepackage{braket}

\usepackage{xcolor}	
\usepackage{newunicodechar}

\usepackage{hyperref}

\newcommand{\wn}{cm$^{-1}$}

\begin{document}
\preprint{APS/123-QED}

\title{Rotational spectra and de-perturbation analysis for ground state ytterbium oxide, YbO}
  
\author{Richard J. Mawhorter}
\affiliation{Department of Physics and Astronomy, Pomona College, Claremont, CA 91711, USA}
 \email{rjm04747@pomona.edu}
 \author{Sean Jackson} \thanks{Graduated Pomona College, 2023}
 \author{Charles Brainin}
\affiliation{Department of Physics and Astronomy, Pomona College, Claremont, CA 91711, USA}
\author{Trevor J. Sears}
 \affiliation{Chemistry Department, Stony Brook University, Stony Brook, NY 11794-3400}
 \email{trevor.sears@stonybrook.edu}
\author{Peter F. Bernath}
 \email{pbernath@odu.edu}
 \affiliation{ Chemistry and Biochemistry Department, Old Dominion University, 4501 Elkhorn Ave., Norfolk, VA 23529}
\author{Philipp Buschmann}
 \affiliation{
 Gottfried Wilhelm Leibniz Universit\"at, Institut f\"ur Physikalische Chemie and Elektrochemie, Hannover, 30167, Germany\\ }
\author{Jens-Uwe Grabow}
 \email{jens-uwe.grabow@pci.uni-hannover.de}
 \affiliation{
 Gottfried Wilhelm Leibniz Universit\"at, Institut f\"ur Physikalische Chemie and Elektrochemie, Hannover, 30167, Germany\\ }

\date{\today} 

\begin{abstract}
A combination difference analysis of Fourier transform microwave spectroscopy (FTMW) measurements of $^{174}$YbO combined with earlier near-infrared chemiluminescence data have resulted in much improved YbO rotational and centrifugal distortion constants.  These have been confirmed in a multi-isotopologue analysis of further FTMW measurements of $^{172}$YbO and $^{176}$YbO and an analysis of nuclear-size-dependent Born-Oppenheimer breakdown effects in the molecule. Multiple excited electronic states overlap with low-lying excited vibrational levels of the electronic ground state leading to irregularities in its vibrational spacings and other previously observed electronic manifolds.  This has been modeled with a matrix of the vibrational levels of the ground and multiple interacting states assuming Morse potentials and including electronic Hamiltonian matrix elements mixing the states. Vibrational level overlaps were calculated numerically assuming Morse vibrational wavefunctions. The resulting de-perturbed potentials for the states inform continuing experimental and theoretical work on this and other ytterbium-containing molecules.

\end{abstract}

\maketitle


\section{\label{sec:intro1}Introduction}
Due to the window it provides into metal-oxygen bonding, the study of the spectra of lanthanide, or rare earth, oxides has been of considerable interest for more than 60 years.  The spectra are complicated because of a plethora of low-lying electronic states arising from the presence of a partially filled $f-$shell on the metal atom.  Classical spectroscopic studies of multiple diatomic lanthanide oxides\cite{Barrow:1979, Kaledin:1981a} showed that many of the observed electronic states shared vibrational frequencies of around 830 \wn and had similar bond lengths, leading to the conclusion that states derived from a particular partially filled $f-$orbital configuration on the metal have similar properties.  These early experimental studies led to the development of a successful conceptual model based on a ligand-field perturbation of the $f-$shell metal ion configuration by the approach of a singly- or doubly-charged oxygen anion\cite{Field:1982}.  This model has successfully predicted the main features and the energy ordering of the low-lying states in rare-earth diatomics.  

Subsequent visible laser-based spectroscopic studies revealed many more details of the energy level structure for diatomic ytterbium oxide, YbO in particular.  Work by the Field group\cite{Linton:1983, McDonald:1985, McDonald:1990} has been pivotal.  The Yb$^{2+}$ ion has a filled, $4f^{14}$, ground configuration and a $4f^{13}6s^1$ low-lying excited energy configuration. Bonding of the O$^{2-}$ anion with the former was postulated to lead to the ground electronic state, $X^1\Sigma^+$, of the molecule, while its bonding with the latter gives rise to the multiple observed low-lying excited molecular states. The Field group's spectroscopic analyses\cite{McDonald:1985, McDonald:1990} mapped out the relative energies of many of the vibronic states below 8000 \wn.  Most importantly, this work established the relative energies of vibrational levels in the ground state where the $4f^{14}$ configuration results in a lower vibrational frequency of 680 \wn compared to some of those in the higher, $f-$hole states, where the vibrational frequency is typically about 830 \wn  ~indicating a stronger metal-oxygen bond.  

This simple picture was belied in 1997 by the measurement\cite{Steimle:1997YbODipole} of 5.89 debye for the dipole moment of the YbO ground state, indicating an admixture of Yb $4f^{14}$ and $4f^{13}6s^1$ configurations.  While earlier computational work did not reproduce this measured dipole moment, a relativistic density functional theory (DFT) study published the following year by Liu, Dolg, and Li\cite{Liu:1998} reported a value of 5.48 debye for a bond length ($r_e$) of 1.865 \AA. Corrected using their $d\mu/dr$ value\cite{Liu:1998} to the actual equilibrium bond length of 1.808 \AA, this results in an estimated 5.07 debye, closer to the value expected for a pure $f^{-1} \sigma^1$ state.\cite{Steimle:1997YbODipole}  Illustrating these difficulties, subsequent DFT work a decade later\cite{Wu:2007DFTofLaOxides} found a dipole moment of 7.5 debye. This was part of a study of the complete set of lanthanum oxides. In this large set, YbO presents a limiting case with the lowest vibrational frequency and among the largest bond lengths and dipole moments. Between these studies a broader study of selected lanthanum diatomics including YbO  using relativistic pseudopotentials and large valence basis sets was published by Cao, Liu, and Dolg.\cite{Xiaoyan2002LaX}  They note that ``the electronic structure of YbO is still an open problem'', a statement which continues to have validity almost 25 years later.

In particular, YbO presents additional complications over and above those found in YbH and YbF, for example, where only singly-charged ion states are important\cite{Liu:1998}.  The difficulties experienced in calculating the ground state wavefunction for YbO derive from the substantial mixing between the filled $f-$shell and $f^{-1}\sigma^1$ Yb$^{2+}$ configurations. For example, Wu et al.\cite{Wu:2007DFTofLaOxides} found an effective 4$f$ population of 13.83, substantial 5$d$ character, and a net metal charge of 1.47 for the ground state wavefunction.  Experimentally, the difference in energy between the ground $^1\Sigma^+$ state and the first excited $0^-$ state is\cite{McDonald:1990} 910 \wn. Published computational work  has had difficulty reproducing this result, let alone the relative positions of the multiple low-lying electronic states\cite{Xu:2009} embedded in the low-lying vibrational levels of the ground and first excited electronic state. 

While the optical laser spectroscopic studies of YbO in oven- or ablation-type sources described above have enabled characterization of the lower electronic states, the upper states involved in these spectra derive from excited $4f^{13}5d^1$ and $4f^{13}6p^1$ configurations of the Yb$^{2+}$ ion as well as states derived from singly-charged ionic configurations.  Their detailed structure has mostly eluded spectroscopic analysis as a result of extensive level perturbations.  More recently, Melville et al.\cite{Melville:2003} measured some near-infrared emission spectra that provided more precise information on the vibrationless ground state structure and some additional information on the character of three excited states lying near 11 000 \wn. 

Current interest in the structure of these molecules focuses on their sensitivity to parity non-conservation (PNC) effects. Ytterbium diatomics and small polyatomics in particular offer a heavy nucleus where these effects are larger and the filled f-shell provides the favorable laser cooling properties of the alkaline earth elements.  The complications of YbO make it less attractive for this purpose, but understanding the complications can greatly aid ongoing and proposed experiments in the YbX family, including YbF \cite{Tarbutt:PRL2026spin}, YbCr \cite{ciamei2026ultracold}, Yb(Cu,Ag,Au),\cite{YbCoinTheory:Poletetal2024} YbOH,\cite{Pilgram:YbOH2021JCP, Takahashi2026:EngineeredMolecularClock} YbNH$_2$,\cite{Vadachkoria:YbNH22025} and YbCaF.\cite{Caldwell:DAMOP2026YbCaF}  

Bare atomic or ionic Yb, with 7 stable isotopes, including two nuclei with non-zero nuclear spins and one with a huge nuclear electric quadrupole moment, also have a lot to offer the field.  Examples include the well-known $^{171}$Yb$^+$ clock transition,\cite{Safronova2018} an interesting Yb$^+$ King plot,\cite{VuleticYb+:2022} and new nuclear radii and magnetic moment results.\cite{Kawasaki2024isotope} Nonetheless, even for the bare atoms and ions, multiple computational difficulties must be overcome.\cite{Safronova2018}

Among other lanthanum oxides, the very recent study of the hyperfine structure of dysprosium oxide, DyO,\cite{Lasner:2026a} demonstrates sensitivity to the nuclear Schiff moment. Gadolinium oxide, GdO,\cite{Kaledin:1994} shares some properties with YbO due to the stability of its half-filled $f-$shell.  The Gd$^{2+}$O$^{2-}$ ionic configurations have mixed $4f^{7}6s^1$ and $4f^{7}6p^1$ character. The half-full $f-$shell confers extra stability compared to most lanthanide oxides leading to state energies that resemble Yb$^{2+}$O$^{2-}$ with its stable full $f-$shell. Furthermore, long-running studies of actinide thorium oxide, ThO, have only recently been surpassed in providing the lowest upper limit on the electric dipole moment of the electron (eEDM)\cite{ACME-I, ACME-II} with ongoing work in progress.\cite{Wu:2020}  
   
Studies of other heavy metal oxide molecules provide contrast and insight.  Detailed microwave spectroscopy studies of hafnium oxide, HfO, \cite{Suenram:1990, Lesarri:2002} provided extremely precise electric nuclear quadrupole eQq splittings for calibrating isoelectronic HfF$^+$ eEDM measurements.\cite{Roussy:2023HfFpluseEDM}  Lead oxide, PbO, was an early candidate for eEDM studies\cite{Petrov:2005PbF, HamiltonPbO:2009}  and a high resolution Fourier transform microwave (FTMW) PbO study sheds light on the electron density at the lead nucleus\cite{Serafin2007}  which contributes to a component of the Born-Oppenheimer breakdown (BOB) effect.\cite{Almoukhalalati2016} Furthermore, a series of studies by a Japanese collaboration\cite{Enomoto:2024PbOStates, Enomoto:2026} has recently culminated in a detailed de-perturbation analysis of multiple, mutually perturbing, low-lying electronic states.  Finally, a thorough rotational spectroscopy study of the BiO radical\cite{Cohen:2006BiO} aided in several experimental studies of isoelectronic lead fluoride, PbF,\cite{Jackson:2024PbFX1X2, Mawhorter2011, Lukas2011} which turns out to offer better opportunities for eEDM and other PNC-related studies than PbO. \cite{Borschevsky2013, Flambaum:2013}

In this paper we report multiple measurements of the primary J = 1$\leftarrow$0 rotational transition for each of the three most abundant even mass Yb isotopologues of YbO. These data have been combined with combination difference data from the near-IR study of Melville et al.\cite{Melville:2003} to provide significantly improved estimates for all the major structural constants describing the $X^1\Sigma^+$ state of the molecule.

We have also made a de-perturbation calculation of multiple mutual local perturbations among vibrational levels of some of the low-lying excited electronic states and excited vibrational levels of the ground electronic state, originally identified by Linton et al.\cite{Linton:1983} and McDonald et al.\cite{McDonald:1985, McDonald:1990}  We have set up a model calculation similar to that described by Zhang et al.\cite{Zhang:2022YbF} for excited states of YbF.  Other oxides have also been the subject of de-perturbation studies including a much earlier study of CO\cite{Klemperer:COdeperturbation} and a very recent study of PbO by Enomoto et al.\cite{Enomoto:2024PbOStates}  In the present work, the electronic states are represented by Morse oscillator potentials and level interactions have been numerically evaluated as products of Morse function overlap integrals and matrix elements of the electronic perturbation Hamiltonian.  The observed energies are the eigenvalues of the perturbation matrix found after optimizing the the Morse potential variables for the interacting states, and the size of the electronic spin-orbit coupling matrix elements.  The results are very satisfactory in terms of reproducing the observed perturbed series of levels and the relative intensities in the published spectra. 
  
\section{\label{sec:expt}Experiment}

We have made multiple measurements of the primary N = J = 1$\leftarrow$0 rotational transition for each of the 3 relatively abundant even mass Yb  ($^{172}$Yb(21.9\%), $^{174}$Yb(31.8\%) and $^{176}$Yb(12.7\%)) isotopologues of Yb$^{16}$O. The transition occurs near 21 GHz and given the maximum spectrometer frequency of 26.5 GHz it was not possible to measure the J = 2$\leftarrow$1 and higher transitions at double this frequency and higher. $^{168}$Yb and $^{170}$Yb abundances are only 0.13$\%$ and 3.5$\%$ respectively, while the odd mass isotopes have non-zero nuclear spins causing dilution of the spectral line intensities. Likewise, $^{17}$O and $^{18}$O abundances are only 0.038$\%$ and 0.205$\%$ respectively, so only YbO molecules containing $^{16}$O nuclei are considered here.

The Fourier transform microwave (FTMW) spectrometer used has been described previously. \cite{GrabowCOBRA1996, GrabowCOBRA2005} Enough background oxygen atoms are present in the nozzle area from residual water in the argon expansion gas that YbO could be synthesized \textit{in situ} by laser vaporization of a sample of natural abundance Yb metal.  YbO molecules are formed in the laser induced plasma by reaction with residual oxygen and from oxidized Yb metal surfaces and then entrained in the stream of high pressure argon gas prior to a supersonic free-jet expansion coaxially to the axis of a Fabry-Perot-type resonator. Frequencies were determined by the Fourier transform of the free induction decay detected by an antenna critically coupled to the resonator tuned to the transition frequency.  This enables sub-kHz resolution for transitions with good signal-to-noise. 

The different even mass Yb isotopologues possess predictably different transition frequencies and spectral assignments were straightforward.  The requirement that the three different isotopic species data are consistent in the overall multi-isotope fit provides very strong limits against possible mis-assignments and provides definite identification of the individual isotopologues. 

\section{\label{sec:res}Results \& Discussion}

\subsection{\label{subsec:const}Rotational Constants}

Nine separate measurements of FTMW rotational J = 1$\leftarrow$0 frequencies have been carried out for the three most abundant even mass YbO isotopologues, i.e. two for $^{174}$YbO near 20992 MHz, three for $^{172}$YbO near 21013 MHz, and four for $^{176}$YbO near 20972 MHz. These transitions and a multiple-isotopologue fit using the program SPFIT\cite{Novick2016} are shown in Table~\ref{tab:FTMWlines+fit}, where the Y$_{01}$ and Y$_{02}$ Dunham parameters correspond to the normal rotational, B, and centrifugal distortion, $-$D, constants. Table~\ref{tab:FTMWlines+fit} also includes the simple fit correlation matrix.

\begin{figure*}[t!]
        \includegraphics[height=5.5cm]{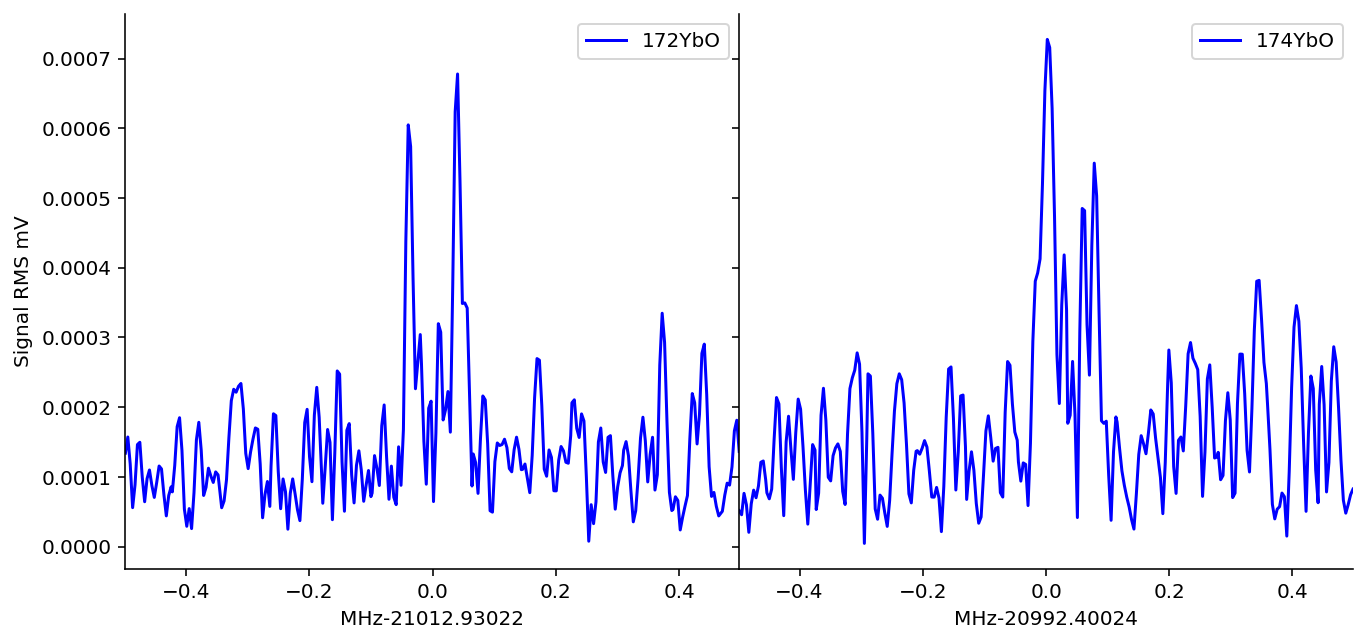}
    \caption{\label{fig:FTMWData}\footnotesize{Examples of FTMW transitions in $^{172}$YbO and $^{174}$YbO. The resonance signal appears twice due to the Doppler shift of the expanding molecular jet's emission with respect to the resonator mode's propagation}. The rest frequency is the arithmetic mean of the two frequencies. The experimental conditions were optimized for each transition.}
\end{figure*}

Spectra for $^{174}$YbO and $^{172}$YbO are shown in Figure \ref{fig:FTMWData}.  A signal to noise ratio of about 3:1 normally indicates that the uncertainty estimate of 1 kHz for these measurements is appropriate.  The quoted parameter uncertainties\cite{Novick2016} in Table~\ref{tab:FTMWlines+fit} for the reference isotopologue $^{174}$Yb$^{16}$O reflect this estimate even though the unitless RMS fit error\cite{Novick2016} of around 0.25 indicates 1-$\sigma$ agreement that is about a factor of 4 better (and also that a higher order Y$_{03}$ (or H) constant is not necessary). This is borne out by the very small observed $-$ calculated frequency differences, which display randomly balanced positive and negative values.  The resulting fit parameters are also not excessively correlated.

\begin{table}[t]
\caption{Multiple-isotope fitted parameters, correlation matrix, and observed FTMW N $=$ 1$\leftarrow$0 transitions in MHz for the three
most abundant even isotopologues of Yb$^{16}$O.  Standard 1$\sigma$ uncertainties are given in
parentheses and the dimensionless RMS deviation of the fit is 0.253.\cite{Novick2016}.
Y$_{01}$ ($=$~B) and Y$_{02}$ ($=$~-D) values in \lbrack ~\rbrack ~are fixed to the corresponding $^{174}$Yb$^{16}$O
reference and are scaled by $1/\mu$ and $1/\mu^2$ respectively.
}
\label{tab:FTMWlines+fit}
\begin{ruledtabular}
\begin{tabular}{cccccc}

\multicolumn{6}{l}{\textbf{FTMW Transitions of YbO (MHz)}} \\
Yb Isotope &	$N'$&	$N''$& Measured Frequency &Obs.-Calc.&	Uncertainty \\

174& 1&	0&	20992.4372&  0.0000& 0.0010 \\

174 & 1&  0&  20992.4373& 0.0000 & 0.0010 \\     
172& 1& 0& 21012.9293& 0.0000 & 0.0010 \\               
172 & 1 & 0 & 21012.9288& -0.0004& 0.0010 \\               
172 & 1 & 0 & 21012.9296 & 0.0004& 0.0010 \\               
176 & 1 & 0 & 20972.3965 &-0.0001& 0.0010\\               
176 & 1 & 0 & 20972.3970 & 0.0004& 0.0010 \\              
176 & 1 & 0 & 20972.3964 &-0.0001& 0.0010 \\              
176 & 1 & 0 & 20972.3965 &-0.0001& 0.0010 \\
\hline
\hline
\multicolumn{4}{l}{\textbf{Fit Parameters}} \\
Yb Isotope & Parameter & Value (MHz) & Isotopic Scaling Ratio & & \\
174 & Y$_{01}$ & 10496.23650(35) & & & \\
172 &  & [10506.53120(35)] &$\mu^{-1}$ & & \\
176 &  & [10486.17003(35)] &$\mu^{-1}$ & & \\
& & & & & \\
172 & Y$_{01}$ BOB & -0.04870(46) & & & \\
176 &  & 0.04609(43) & & & \\
& & & & & \\
174 & Y$_{02}$ & -0.008931\footnotemark[1] & & & \\
172 &  & [-0.008949] &$\mu^{-2}$ & & \\
176 &  & [-0.008914] &$\mu^{-2}$ & & \\
\hline
\hline
\multicolumn{4}{l}{\textbf{Correlation Matrix}} \\
 & Y$_{01}$ & $^{172}$BOB  & $^{176}$BOB   \\
Y$_{01}$ & 1.0000 & & \\
$^{172}$BOB & -0.7744 &  1.0000 & \\
$^{176}$BOB & -0.8158 &  0.6318 & 1.0000 \\
\end{tabular}
\end{ruledtabular}
\footnotetext[1]{The Y$_{02}$ ($-$D) value is fixed to the FTMW + NIR combination difference D-value
determined herein.}
\end{table}

This is an exceedingly consistent fit but the observed $^{174}$YbO transition frequency was initially surprising because it was found far below the search starting point based on values from previous studies,\cite{Melville:2003, Linton:1983} where corresponding $Y_{01}$ (B) values several $\sigma$ higher were reported.

In order to resolve the apparent disagreement, we combined the present FTMW data with combination differences derived from the Melville System 1 and System 2 data\cite{Melville:2003} which includes J-values between 1 and 48 and an estimated measurement uncertainty of 0.005 \wn (= 150 MHz). The resulting rotational and centrifugal distortion constants are shown in Table~\ref{tab:B+D} in both \wn and MHz units. The combined fitting shows our measured data is consistent with the larger NIR data sets, and the results highlight the extra information resulting from combining these complementary measurements.  The more robust B and D values shown in Table~\ref{tab:B+D} exhibit reductions of factors of 900 and 2.5 in their respective uncertainties.  The new combined B value of 10496.238(1) MHz is essentially clamped to that derived from the FTMW measurements due to their high statistical weighting. The now better-determined centrifugal distortion constant (D) of 8.93(13) kHz is much smaller than the Linton value\cite{Linton:1983} and ~30$\%$ less than reported in the Melville et al.\cite{Melville:2003} analysis. Since it is determined by the high-J NIR data, it was subsequently fixed in the FTMW multi-isotope fit, permitting better determination of the large Born-Oppenheimer breakdown terms expected because of the numerous low-lying excited states in YbO. A check of the D value via the Kratzer-Pekeris relationship\cite{GordyandCook} finds a value of 11.0 kHz compared to the value determined here of 8.93(13) kHz. Since the Kratzer-Pekeris derivation assumes a Morse potential, this mismatch provides another perspective on the complexity of the YbO ground state.

The availability of both the $B_0$ and $B_1$ rotational constants for the $v = 0$ and $v = 1$ levels of the ground electronic state in the NIR study of Melville et al.\cite{Melville:2003} enables determination of an equilibrium rotational constant $B_e$ and the associated rotation-vibration interaction constant $\alpha_e$.  The corresponding  $r_0, r_1$, and $r_e$ bond distances have also been determined, and are shown in Table~\ref{tab:Be+re}, along with an another set using the much more precise FTMW $B_0$ value and the NIR $\alpha_e$.  This nudges the $r_e$ value from 1.80706(14) \AA  ~to 1.80760(14) \AA ~and we will use the 1.808 \AA ~bond distance value in our subsequent work.

\begin{table}[h]
\caption{$^{174}$YbO Rotational parameters in both \wn ~and MHz compared with literature values and with the added combination difference method employed for the data in \cite{Melville:2003} shown here, both with and without the high resolution FTMW $^{174}$YbO data point. The new B value from the FTMW multiple-isotopologue fit of 9 transitions from the 3 even $^{172}$YbO, $^{174}$YbO, and $^{176}$YbO isotopologues is also provided for comparison. $1\sigma$ 
uncertainties are in parentheses.}
\label{tab:B+D}
\begin{ruledtabular}
\begin{tabular}{ccccc}
\textbf{$^{174}$YbO Rotational Constants} &	\textbf{B} (\wn) &	\textbf{D} ($10^{-7}$\wn) & \textbf{B} (MHz) & \textbf{D} (MHz) \\

FTMW only multi-isotop. fit\footnotemark[1] & & & 10496.23650(35)	& 0.008931 (\textit{fixed}) \\

FTMW+NIR comb. diff.\footnotemark[1] & 0.3501168(70)& 2.978(43) & 10496.238(1)	& 0.008931(130) \\

NIR System 1 \& 2 comb. diff.\footnotemark[1] & 0.350314(32) & 3.85(10) & 10502.15(95)	& 0.0115(3) \\

NIR Systems 1 \& 2 (2003) & 0.350326(27) & 3.91(9) & 10502.51(81)	& 0.0117(3) \\

Laser Spectroscopy (1983) & 0.350220(40) & 7.8(1.7) & 10499.33(120)  & 0.0234(51) \\
\end{tabular}
\end{ruledtabular}
\footnotetext[1]{This work}
\end{table}

\begin{table}[h]
\caption{$^{174}$Yb$^{16}$O Equilibrium Rotational Constants and Bond Distances.  Using an $\alpha$ parameter derived from the NIR study of Melville, et al. \cite{Melville:2003},
equilibrium rotational constants B$_e$ and the corresponding bond distances are are derived using NIR B$_0$ and B$_1$ values as well the FTMW B$_0$ from this study.  Unless otherwise specified, rotational constants B are in units of \wn and bond distances r are in \AA. $1\sigma$ 
uncertainties are in parentheses.}
\label{tab:Be+re}
\begin{ruledtabular}
\begin{tabular}{ccc}

\textbf{Parameter (Data Set)} &	\textbf{\wn} &	\textbf{\AA} \\

B$_0$/$r_0$ (NIR)	& 0.350326(27) &1.812483(70) \\

B$_1$/$r_1$ (NIR) & 0.346116(48) &1.823473(126) \\     
B$_e$/$r_e$ (NIR) & 0.352431(55) &1.807062(140)\\  

$\alpha_e$ (NIR) &0.004210(55) & \\           & & \\
Y$_{01}$ $\sim$ B$_0$ (FTMW only, \wn ~below)& 10496.23650(35) MHz& \\
B$_0$/$r_0$ (FTMW only)& 0.350116763(12) & 1.81302430(3) \\
B$_1$/$r_1$ (FTMW+NIR $\alpha$)&   0.345907(55)&	1.824024(150) \\
B$_e$/$r_e$ (FTMW+NIR $\alpha$) &	0.352222(55)&	1.807599(140)\\

\end{tabular}
\end{ruledtabular}
\end{table}

\subsection{\label{subsec:BOB}Born-Oppenheimer Breakdown Analysis}

As noted above, the multi-isotopologue FTMW data set was also analyzed on its own to check for consistency.  In this process, the ratios of the parameter values for different isotopologues relative to the reference isotopologue (here, the most abundant $^{174}$YbO) were fixed according to the appropriate reciprocal powers of the reduced mass $\mu$\cite{Drouin2001} while only $Y_{01}$ and $Y_{02}$ values for the reference isotopologue were varied. For example, the even isotopologue rotational constants B = $Y_{01}$ = $h/(8 \pi \mu r^2)$, was scaled as 1/$\mu$.

This analysis resulted in very large Born-Oppenheimer breakdown (BOB) terms of 45-50 kHz for $^{172}$YbO and $^{176}$YbO with respect to the reference $^{174}$YbO isotopologue.  The Born-Oppenheimer approximation assumes that (1) the nucleus is infinitely heavy relative to the surrounding electrons, (2) that the nucleus is a point particle with zero radius, and (3) that the electronic state in question is isolated.  The first assumption is less valid for light nuclei and so is of little consequence here. On the other hand, nuclear radii grow as more nucleons are added and the Yb nuclear radii are nearly monotonically larger as the number of neutrons increases here from mass 172 to 176. 

Our recent studies of BaF\cite{Preston2026globalBaF} and PbF\cite{Jackson:2024PbFX1X2} where overlapping states are not an issue provide some helpful perspectives for estimating the size of BOB terms stemming from assumptions (1) and (2) above in the intermediate case of YbO.    For BaF the BOB data from $^{138}$BaF to $^{134}$BaF follow the zigzag pattern of the barium nuclear radii and the overall BOB effect is about 1 kHz, with about 1/3 of the total coming from assumption (1), the mass dependent effect.  For PbF the overall BOB effect is much larger at about 8 kHz and the smaller mass dependent term (1) is expected to be less than a few percent of the total. This is borne out by the excellent fit to the BOB residuals using only the nuclear radius terms (2) going from $^{208}$PbF to $^{204}$PbF.  Hence for YbO we may generally anticipate a nuclear radius effect of a few to several kHz and a tiny mass dependent effect of a fraction of a kHz.

Thus perturbing electronic states, violating assumption (3), are required to cause the observed, large, BOB effects in YbO, and this is exactly what we have here, even for the ground state.  To get a sense of scale, TeSe is of similar mass and exhibits BOB terms of 20 kHz.\cite{Banser2006TeSeJMolSpec}   Although the influence of the overlapping states’ perturbations makes it difficult to separate the nuclear radius size field shift and mass dependent BOB terms, a closer look of the more important nuclear size terms is worthwhile.
Taking a specific look at the nuclear radii, the values of the 2013 Angeli encyclopedic table\cite{Angeli2013} for the corresponding $\delta<r^2>$ shifts grow by 4$\%$ more between Yb 172 and 174 nuclei than between $^{174}$Yb and $^{176}$Yb  The Landolt-B\"ornstein reference for Yb\cite{Landolt-Bornstein} indicates a similarly larger 3.5$\%$ increase (with larger error limits). More recently, Kawasaki et al.\cite{Kawasaki2024isotope} found a very similar 3.7$\%$ larger increase for the 172-174 masses compared to the change between the $^{174}$Yb and $^{176}$Yb nuclei. The mass dependent BOB terms are also larger for lighter mass nuclei so they lean very slightly in this direction as well.

In the YbO data presented here, the negative BOB shift to mass 172 of -48.7(5) kHz from the mass 174 reference is 6$\%$ larger than the observed positive shift of 46.1(4) kHz from mass 174 to mass 176. These trends are consistent with the nuclear radii shifts, but the fact that we expect the nuclear radius field shift itself to be about 1/10 or less of the total 45 kHz precludes making a detailed comparison.  Nevertheless, these consistencies lend credence to the overall BOB shifts we determine, and their very large size aligns with the fact that the ground state of YbO suffers perturbations from multiple low-lying excited electronic states.

\subsection{\label{subsec:overlap}X--B--D State Overlap} 

Spectroscopic work by Linton \cite{Linton:1983} and McDonald\cite{McDonald:1990} established the energies of multiple low-lying excited electronic states in YbO. The observed vibrational energies of the ground electronic state exhibit significant deviations from the lower-\(v\) anharmonic trend around $v=4$. Previous analyses attributed this anomalous behavior to perturbative interactions with nearly degenerate vibrational levels belonging to electronically excited states. In particular, at around 3000 cm$^{-1}$, the \(v=4\) of the ground, $X^1\Sigma^+$, state is nearly degenerate with the \(v=2\) level of the first excited $\Omega = 1$ state, referred to as the $B_1-$state below, and the \(v=0\) level of a second $\Omega = 1$, labeled $\Omega = 1^{\prime}$ by McDonald et al.,\cite{McDonald:1990} and referred to as the $D_1-$state below. Both of these states are predominantly derived from the $f^{-1}$ configuration.  While electronic spin-orbit mixing terms in the molecular Hamiltonian are often assumed to be diagonal in the projection quantum number $\Omega$, operators mixing states differing in $\Omega$ exist in the microscopic Hamiltonian.\cite{Brown2003}  Their effects on the energies derived from an effective Hamiltonian are usually small for systems with isolated electronic states and are conventionally folded into the effective state origins, $T_e$.  This situation does not hold for YbO, and we allow for vibrational level-mediated state coupling terms in the model described below. We use the shorthand $H^{\prime}_{XB_1}$ for example to refer to the perturbation matrix element between the $X$ and $B_1$ states below.

\begin{table}[t]
\caption{Experimental vibrational transition frequencies and electronic state origins ($\text{cm}^{-1}$) utilized for model validation.}
\label{tab:Linton&McDonald Data}
\begin{ruledtabular}
\begin{tabular}{cc}
Transition ($v' \rightarrow v''$) & Frequency / Origin \\
\hline
\multicolumn{2}{l}{\textbf{Ground State $X$ Transitions}\footnotemark[1]} \\
1 $\rightarrow$ 0   & 683(2)   \\
2 $\rightarrow$ 1   & 663(1)   \\
3 $\rightarrow$ 2   & 651(1)   \\
4 $\rightarrow$ 3   & 633(4)   \\
5 $\rightarrow$ 4   & 627(1)   \\
6 $\rightarrow$ 5   & 617(2)   \\
7 $\rightarrow$ 6   & 611(3)   \\
8 $\rightarrow$ 7   & 605(3)   \\
9 $\rightarrow$ 8   & 555(10)  \\
10 $\rightarrow$ 9  & 609(10)  \\
\hline
\multicolumn{2}{l}{\textbf{Excited State $B_1$ Transitions}\footnotemark[2]} \\
1 $\rightarrow$ 0   & 822(6)      \\
2 $\rightarrow$ 1   & 830(3)      \\
3 $\rightarrow$ 2   & 831(6)      \\
\hline
\multicolumn{2}{l}{\textbf{Electronic State Origins ($T_0$)}} \\
Ground State $X$    & 0        \\
Excited State $B_1$ & 1015(10) \\
Excited State $D_1$\footnotemark[3] & 2702(10) \\
\end{tabular}
\end{ruledtabular}
\footnotetext[1]{Values for the $X$ state represent the average value of Linton and McDonald where multiple values are available, Linton if only one, and McDonald data otherwise.}
\footnotetext[2]{Only transitions up to $v'=3$ are experimentally known for the $B_1$ state.}
\footnotetext[3]{Only the $v=0$ state origin ($T_0$) is experimentally known for the $D_1$ state.}
\end{table}

Our objective is to determine de-perturbed potential energy curves for the $X$, $B_1$, and $D_1$ electronic states using a coupled-state model similar to the recent studies of YbF\cite{Zhang:2022YbF} and PbO.\cite{Enomoto:2026}  In the present approach, we consider a 3-state mixing using Morse oscillator wavefunctions to model the vibrational overlap to account for the strong anharmonicity in the YbO ground state. In the model, developed in Python,\cite{Hill2023:python} the vibrational levels associated with electronic state \(s\in\{X,B_1,D_1\}\) labeled as \(v_s\) correspond to energies
\begin{equation}
E_s(v_s)=T_{e,s}
+\omega_{e,s}\left(v_s+\tfrac12\right)
-\omega_{e,s}x_{e,s}\left(v_s+\tfrac12\right)^2
+\omega_{e,s}y_{e,s}\left(v_s+\tfrac12\right)^3.
\end{equation}
Here, \(T_e\) denotes the electronic term energy, \(\omega_e\) is the harmonic vibrational constant, and \(\omega_ex_e\) and \(\omega_ey_e\) represent first and second anharmonic corrections, respectively. Inclusion of the cubic ($\omega_ey_e$) term was found to be necessary to adequately reproduce the positions of the higher vibrational levels of the X-state manifold although Morse wavefunctions (which refer to an anharmonic Hamiltonian including only the quadratic term) are assumed.

Representing the interacting $X, B_1,\ \text{and}\ D_1$ states with this model requires 5 parameters ($T_e,\ \omega_e,\ \omega_ex_e,\ \omega_ey_e,\ r_e$) per state totaling 15 independent parameters before mixing is considered. Since the experimental data are restricted, we reduced the number of anharmonic terms for the excited states as detailed below.  To account for mixing, we include the electronic coupling terms between pairs of states $H_{E_{B_1X}}, \ H_{E_{XD_1}}, H_{E_{B_1D_1}}$ where $H_{E_{s_is_j}} = \braket{s_i|H'|s_j}$. Under the Born--Oppenheimer approximation, the electronic and vibrational degrees of freedom are assumed to separate such that $\braket{s_i,v_{s_i}|H'|s_j,v_{s_j}}=\braket{s_i|H'|s_j}\braket{v_{s_i}|v_{s_j}}=H_{E_{s_is_j}}F_{s_i s_j},$
where
$\lvert F_{s_is_j}\rvert^2 = \lvert \braket{v_{s_i}|v_{s_j}}\rvert^2$
is the vibrational overlap integral square, or Franck--Condon Factor (FCF), between the two states. The effective Hamiltonian we construct in the $\ket{s,v_s}$ basis is of the form 
\begin{equation}
    H_{\text{eff}} = \sum_{s,v} E_s(v)\ket{s,v_s}\bra{s,v_s} + \sum_{s\neq s',v,v'}H_EF_{v,v'}\ket{s,v_s}\bra{s',v'_{s'}}.
    \label{Heff}
\end{equation}In total, this would yield 18 free parameters, which is greater than the experimental data available as shown in Table \ref{tab:Linton&McDonald Data}. 

To model the limited perturbed level data, a reduction in the number of free parameters is required. We set $T_{e,X} = 0$ and reference the term energies of the $B_1$ and $D_1$ to this value. In the same way, our analysis need only consider the differences in $r_e$, that is, $\Delta r_{B_1}$ and $\Delta r_{D_1}$ between the ground and excited states rather than the absolute values for each state. We use the NLxc theory values of $\Delta r_{B_1}$ and $\Delta r_{D_1}$ from Liu et al.\cite{Liu:1998} ($\Delta r_{B_1} =$ -0.042 \AA ~and $\Delta r_{D_1} =$ -0.022 \AA) which we fix in the proceeding analysis; we note that taking the average of the four theory levels presented by Liu et al. yields very similar results. The $\omega_ey_e$ term is only required in the $X$ state to account for its anharmonicity, it is therefore fixed to 0 in both the $B_1$ and $D_1$ states. Furthermore, we note the importance of the $D_1$ state in resolving the perturbations present in the $X(v=4)$ level; the observed perturbations cannot be accounted for by a single perturbing level.  However, since only the origin of the $D_1$ state has been observed, we cannot independently determine its potential. Accordingly, we fix $\omega_{e,{D_1}}$ in the ratio predicted by the average of the theoretical results.\cite{Liu:1998} Lastly, we assume that it has the same anharmonicity as the $B_1$ state as they are both predominantly $f-$hole configuration derived, and we therefore fix $\omega_ex_{e,D_1} = \omega_ex_{e,B_1}$, accordingly. In this way, the the 18 parameters originally in the model used are reduced to 10:
$\omega_{e,X},\ \omega_ex_{e,X},\ \omega_ey_{e,X}, T_{e,B_1},\ \omega_{e,B_1},\ \omega_ex_{e,B_1},\ T_{e,D_1},\ H_{E_{B_1X}}, \ H_{E_{XD_1}}, H_{E_{B_1D_1}}$, where the subscript denotes which state the parameter belongs to. To the extent that \eqref{Heff} is an accurate model, these 10 parameters can be determined by fitting the eigenvalues of the interaction matrix to the experimentally measured energy levels for each of the three states. The Hamiltonian matrix was set up including vibrational basis functions $0 \le v_{X} \le 12$, $0 \le v_{B_1} \le 9$ and $0 \le v_{D_1} \le 6$.

Initial parameter estimates were obtained by a combination of independently fitting each electronic state prior to inclusion of interstate coupling as well as theoretical calculations from Liu et al.\cite{Liu:1998}. It was initially assumed the ground state vibrational levels $v= 0,\ 1,\ 2,\ \text{and}\ 3$ are unperturbed without near-degenerate energy levels, so initial X-state Morse vibrational parameters were fitted to reproduce these spacings yielding the parameters: $\omega_{e,X} = 696\ \text{cm}^{-1},\ \omega_ex_{e,X} = 8.84\ \text{cm}^{-1},\ \omega_ey_{e,X} = 0.25\ \text{cm}^{-1}$. Apparent negative anharmonicity present in the $B_1$-state prevents a physically sensible fitting so we base the initial guesses on values reported from\cite{McDonald:1990}\cite{Linton:1983}: $T_{e,B_1} = 1016\ \text{cm}^{-1},\ T_{e,D_1} = 2702\ \text{cm}^{-1},\  \omega_{e,{B_1}}=825\ \text{cm}^{-1},\ \omega_{e,{D_1}}=810\ \text{cm}^{-1}$.  Additionally we set $\ \omega_ex_{e,{B_1}} = \omega_ex_{e,{D_1}}=4\ \text{cm}^{-1}$ as physically plausible initial guesses. With no predictions available for the electronic mixing $H_E$, we make an initial guess of $10\ \text{cm}^{-1}$ for $H_{E_{X,B_1}}$ and $H_{E_{X,D_1}}$ and a larger value of $30\ \text{cm}^{-1}$ for $H_{E_{D_1,B_1}}$ since the $B_1$ and $D_1$ states have the same $\Omega$.

The vibrational wavefunctions, $\psi_v(r)$, for each state were modeled analytically using the Morse oscillator framework. For a given electronic state ($X$, $B_1$, and $D_1$) and vibrational quantum number ($v$), the spatial profile of the wavefunction is formulated and modeled by the Morse SciPy project from Hill.\cite{Hill2023:python} We calculate overlap integrals between two distinct electronic states, $A$ and $B$, directly via numerical integration over a shared $r-$coordinate grid thus eliminating the need for a variable Jacobian:
\begin{equation}
    \langle \psi_{v'}^A \mid \psi_{v''}^B \rangle = \int_{r_{\text{min}}}^{r_{\text{max}}} \psi_{v'}^A(r) \, \psi_{v''}^B(r) \, dr.
    \label{eq:overlap_integral}
\end{equation}
We evaluated this integral via a numerical trapezoidal integration over a dense mesh of ($N = 4000$) points spanning from $r_{\text{min}} = 1.55\,\text{\AA}$ to $r_{\text{max}} = 2.60\,\text{\AA}$, which covers sufficiently the limits of the wavefunctions.

The eigenvalues of the Hamiltonian matrix were then fit to the experimentally measured vibrational levels of the ground state through \(v=8\) and the measured $B_1$ and $D_1$ state levels through a manual refinement process to reduce the RMSE. We only attempt to fit through $v=8$ because additional heavy perturbations from higher energy states occur above this level\cite{McDonald:1990} which are not accounted for in the model. The resulting optimized parameter set is listed in Table~\ref{tab:fitted_params_vertical}. 

\begin{figure}[h!]
    \centering
    \includegraphics[width=0.45\textwidth]{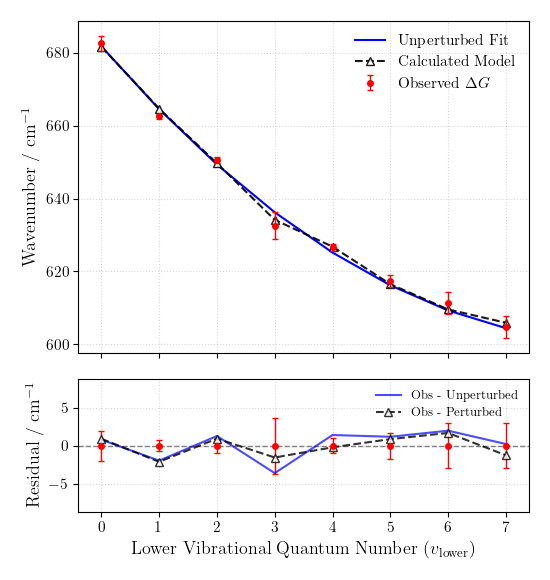}
    \label{fig:Pertgraphic}
\caption{\textbf{De-perturbation analysis of the YbO $X$-state vibrational spacings truncated at $v_{\mathrm{lower}} = 7$ to isolate localized electronic coupling effects.} The top panel compares the experimental data (red circles) against the unperturbed cubic anharmonic oscillator baseline (solid blue line) and the fully diagonalized effective Hamiltonian model (dashed black line with open triangles). The bottom panel illustrates the corresponding residuals ($\text{Observed} - \text{Calculated}$) relative to the experimental zero baseline. Incorporating the explicit state-to-state matrix elements effectively handles the irregularities around $v_{\mathrm{lower}} = 3$, capturing the perturbation-induced shifts to within experimental uncertainty.}\end{figure}

The final model reproduced the measured energy values with a root-mean-square deviation of (8.4~\wn). Notably, the RMSE is two orders of magnitude smaller than the spacing between subsequent energy levels and commensurate with the uncertainty in the experimental data itself. We note that the RMSE is dominated by the $B_1$-state intervals which we believe is caused by the presence of the lowest excited, $\Omega=0^-$, state forcing the $B_1(v=0)$ level higher and an $\Omega=2$ state affecting higher vibrational levels. Neither of these two additional states have been included in this model, as the primary focus was resolving the $X$-state spacing anomalies and there is insufficient experimental data.  We note however that McDonald et al.\cite{McDonald:1990} identified a large $\Omega-$doubling in the $\Omega=1$ (B$_1$) state in question.  They ascribe it to interaction with the $0^-$ state and were able to model it satisfactorily assuming second order perturbation theory involving off-diagonal matrix elements of the rotational Hamiltonian.\cite{Brown2003} We speculate that the $v=0$ level of the B$_1$ state is pushed up in energy through higher order spin-orbit mixing with the $0^-$ state since we see no other explanation for the observed negative anharmonicity in B$_1$.  Since the $0^-$ state is not included in our model, we not expect to have as good agreement for the $B_1$ state level positions. However, Fig.\ref{fig:Pertgraphic} shows the model reproduces the observed perturbations in the ground state levels to well within the measurement uncertainties.

\begin{table}[t]
\caption{Final molecular constants and electronic coupling parameters ($\text{cm}^{-1}$) returned by fitting with the effective Hamiltonian model. Equilibrium bond lengths ($r_e$) are reported in \text{\AA}.}
\label{tab:fitted_params_vertical}
\begin{ruledtabular}
\begin{tabular}{lc}
Parameter & Value \\
\hline
\multicolumn{2}{l}{\textbf{Ground State $X$}} \\
$T_e$          & 0                     \\
$r_e$          & 1.808\footnotemark[1] \\
$\omega_e$     & 701                   \\
$\omega_e x_e$ & 10.2                  \\
$\omega_e y_e$ & 0.34                  \\
\hline
\multicolumn{2}{l}{\textbf{Excited State $B_1$}} \\
$T_e$          & 936                   \\
$r_e$          & 1.765\footnotemark[1] \\
$\omega_e$     & 848                   \\
$\omega_e x_e$ & 6.5                   \\
$\omega_e y_e$ & 0\footnotemark[1]     \\
\hline
\multicolumn{2}{l}{\textbf{Excited State $D_1$}} \\
$T_e$          & 2635                \\
$r_e$          & 1.785\footnotemark[1] \\
$\omega_e$     & 833\footnotemark[2]   \\
$\omega_e x_e$ & 6.5\footnotemark[3]   \\
$\omega_e y_e$ & 0\footnotemark[1]     \\
\hline
\multicolumn{2}{l}{\textbf{Electronic Coupling ($H_{ij}$)}} \\
$\langle X | H_E | B_1 \rangle$     & 20 \\
$\langle X | H_E | D_1 \rangle$     & 20 \\
$\langle B_1 | H_E | D_1 \rangle$ & 50 \\
\end{tabular}
\end{ruledtabular}
\footnotetext[1]{Value held fixed during the optimization procedure.}
\footnotetext[2]{Value held fixed in a constant ratio scaled to the $B_1$ \textit{ab initio} calculation.}
\footnotetext[3]{Value fixed to match the corresponding parameter of the $B_1$ state.}
\end{table}

\subsection{\label{subsec:BR}Spectroscopic Branching Ratios}

To validate the wavefunctions derived in the present work, they were used to compute fluorescence Franck--Condon branching ratios to compare with relative laser-induced fluorescence intensities reported by McDonald \textit{et al.} for the strongly perturbed state accessed at 4518~\AA.~\cite{McDonald:1990} This high-lying state, located at $T_0 \approx 22\,100~\mathrm{cm^{-1}}$ with $\Omega' = 1$, fluoresces to both vibrational levels in the $X\,{}^1\Sigma^+$ ground state and in the lowest $f^{13}s$ ($\Omega = 1$) excited state, with observed intensity ratios of $80{:}10{:}1$ and $30{:}1$, respectively.\cite{McDonald:1990}

Because the 4518~\AA\ excitation band is too strongly perturbed for rotational analysis, its equilibrium bond length cannot be determined directly from spectroscopic constants. The upper-state potential was therefore represented by a Morse oscillator with $r_e = 1.780~\mathrm{\AA}$, $\omega_e = 820~\mathrm{cm^{-1}}$, and $\omega_e x_e = 6.0~\mathrm{cm^{-1}}$. The assumed bond length lies between the experimentally determined values for the $X$ and $B_1$ states, $r_e^X = 1.808~\mathrm{\AA}$ and $r_e^B = 1.765~\mathrm{\AA}$, consistent with the expected intermediate bonding character of the upper state. Furthermore, the calculated branching ratios were found to be largely insensitive to the assumed vibrational frequency, with comparable agreement obtained for $\omega_e$ values spanning $740$--$860~\mathrm{cm^{-1}}$, indicating that the Franck--Condon distribution is governed primarily by the choice of excited state $r_e$.

Franck--Condon factors were calculated for transitions to $v'' = 0$--2 of the $X$ state and $v'' = 0$--1 of the $B_1$ state. Expected relative fluorescence intensities were then obtained using the Einstein $A$ coefficient,
\begin{equation}
I_{v'v''} \propto \nu^{3} q_{v'v''},
\end{equation}
where $\nu$ is the transition frequency and $q_{v'v''}$ is the Franck--Condon factor \cite{Herzberg1950,Hilborn:1982}. The $\nu^3$ dependence is appropriate here because the resolved fluorescence reported by McDonald \textit{et al.}\cite{McDonald:1990} was recorded with a photon-counting detector, for which the measured signal tracks the photon emission rate (the Einstein $A$ coefficient) rather than the radiated power. \cite{Hilborn:1982} We note that because the fluorescence bands compared here span only a narrow range in $\nu$, our qualitative conclusions are essentially unchanged if the $\nu^4$ form appropriate to a power-sensitive detector (Einstein $B$ coefficient) is used instead.

 The resulting calculated branching ratios of $80{:}10.4{:}0.7$ for the $X$-state progression and $30{:}1.1$ for the $B_1$-state progression are in excellent agreement with the observed values of 80:10:1 and 30:1 allowing for expected experimental measurement errors. Given that the molecular constants of the 4518~\AA\ state cannot be determined directly, the simultaneous reproduction of two independent fluorescence channels provides strong support for the adopted upper-state geometry and serves as an independent validation of the wavefunctions derived in the present analysis.

\begin{figure}[htp]
    \centering
    \includegraphics[width=0.45\textwidth]{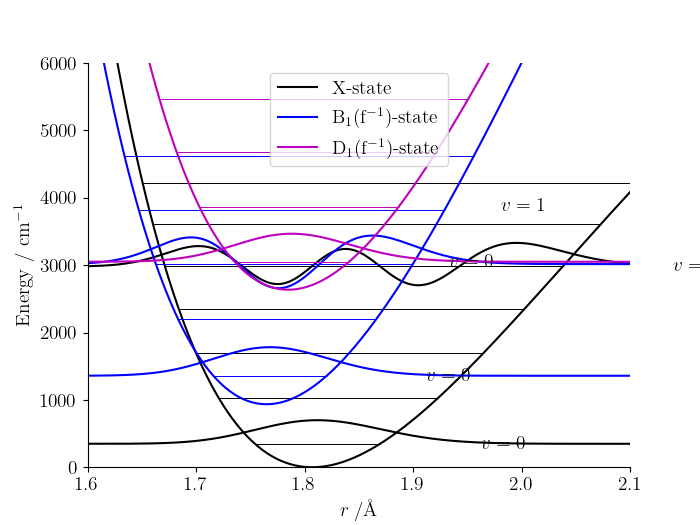}
\caption{\textbf{Calculated potential energy curves ($V(r)$) and selected vibrational wavefunctions for the interacting electronic states of YbO.} The ground state $X$ (black curves), the excited $B_1(f^{-1})$ state (blue curves), and the excited $D_1(f^{-1})$ state (magenta curves) are plotted as a function of the internuclear separation $r$ ($\text{\AA}$). Horizontal lines indicate the calculated vibrational energy levels ($v$) for each respective electronic potential. Representative wavefunctions ($\psi(r)$) are superimposed over their corresponding energy baselines to illustrate  where strong local perturbations and state mixing occur.}
    \label{fig:Wavefns}
\end{figure}

\begin{figure}[h!]
    \centering
    \includegraphics[width=0.45\textwidth]{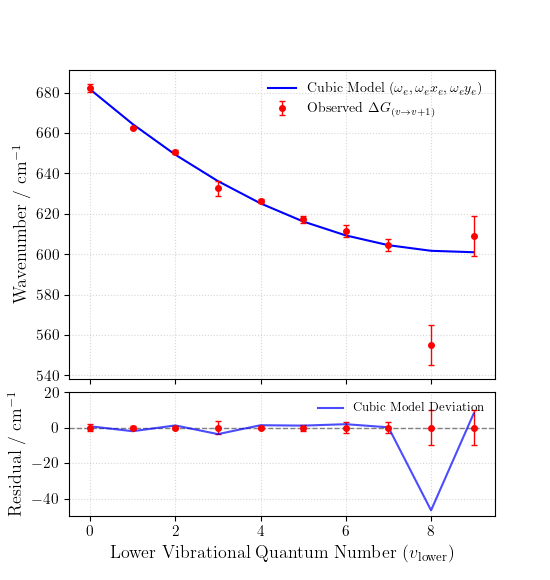}
\caption{\textbf{Experimental and unperturbed vibrational transition spacings ($\Delta G_{v \rightarrow v+1}$) for the ground electronic state of YbO across the full observed range up to $v_{\mathrm{lower}} = 9$.} The top panel displays the experimental spacings (red circles) overlaid with a standard unperturbed cubic polynomial fit based on the equilibrium spectroscopic parameters $\omega_e$, $\omega_ex_e$, and $\omega_ey_e$ (solid blue line). Error bars indicate experimental uncertainties ($1\sigma$). The bottom panel shows the residuals ($\text{Observed} - \text{Calculated}$) centered on the experimental baseline, illustrating the severe breakdown of the unperturbed cubic model at higher vibrational levels due to local perturbations.}\end{figure}

\section{\label{sec:concl}Summary and Conclusions}

In this study high resolution FTMW measurements of the fundamental J = 1$\leftarrow$0 rotational transitions for 3 even ytterbium isotopologues of YbO are presented and analyzed together with existing near-IR measurements.\cite{Melville:2003} Beyond extending the range of isotopes studied, this has resulted in much-improved rotational constants, bond lengths and very large Born-Oppenheimer breakdown terms consistent with the extensive set of low-lying states previously observed.\cite{Linton:1983, McDonald:1990}

A de-perturbation analysis extending up to v = 8 level of the ground electronic state and two other low-lying excited states has also been carried out using Morse oscillator potential functions and results validated by comparison with experimentally observed intensities. The abnormally small $v$=4-3 X state vibrational spacing noted previously\cite{Linton:1983} can be understood by accounting for the near degeneracy of the $v$=2 level of the $B_1$ state and the $v$=0 level of the $D_1$ state with the X state $v$=4 level, and the model also accounts for other observed trends in the measurements. The observed $X-$state level spacings cannot be modeled by a single perturbing state.  This work extends the available techniques for disentangling ubiquitous overlapping f-hole states in ytterbium-containing molecules initiated by harmonic de-perturbation analysis of the A-state of YbF.\cite{Zhang:2022YbF}  That study helped enable an important further understanding of YbF,\cite{PopaPRX2024} and tools like this can likewise assist the numerous studies of diatomic and polyatomic Yb-containing molecules currently underway.

Experimental extensions of this work could include measuring more rotational transitions in I=$\frac{1}{2}$ $^{171}$YbO and possibly extending this to even more complex I=$\frac{5}{2}$ $^{173}$YbO. A survey for observation of the $^{173}$YbO nuclear quadrupole coupling eQq hyperfine structures would require a sensitive broadband microwave spectrometer able to reach 21 GHz and beyond but would provide another experimental signpost beyond the dipole moment\cite{Steimle:1997YbODipole} for the difficult theoretical challenges YbO still poses. This would follow the path taken earlier with open shell $^{171}$YbF\cite{Glassman2014hyperfine} and then $^{173}$YbF \cite{Wang173YbF2019} where good agreement with theory for $^{173}$YbF was achieved.\cite{PastekaYbF2016relativistic} A direct comparison of open shell/closed shell YbF/YbO along these lines would also be potentially interesting, and enlarging the number of isotopes in the multiple-isotopologue fit of the microwave data would provide more detailed BOB information.

\section*{Acknowledgements}

The authors thank Chi Zhang, Lukas Pasteka, Anastasia Borschevsky, and Lan Cheng for helpful discussions. We greatly appreciate the development and availability of the Morse Oscillator Python Class code by, and the helpful discussions with, Christian Hill.  Funding for the Pomona College authors has been provided by a Pomona College Sontag Fellowship as well as the Summer Undergraduate Research Program. R.M. also received support from the German Academic Exchange Service (DAAD) and from the National Science Foundation (NSF grant CHE-2534451). J.-U.G.and P.B. gratefully acknowledge funding by the Deutsche Forschungsgemeinschaft (DFG, German Research Foundation) Project number (461997427).

\pagebreak

\newpage
\bibliography{YbORefs, PbFRefs, PetrovLib}

\end{document}